\documentclass[runningheads]{llncs}
\usepackage{bbding}
\usepackage{xcolor}
\usepackage[pdftex,colorlinks=true,pdfstartview=FitV,linkcolor=black,citecolor=blue,urlcolor=black]{hyperref}

\renewcommand\orcidID[1]{}
\author{Ahmed Najeeb\orcidID{0009-0007-2379-368X} \and
Billy Bob Brumley\orcidID{0000-0001-9160-0463}}
\institute{Rochester Institute of Technology, Rochester, New York, USA\\
\email{\{an4606,bbbics\}@rit.edu}}
\titlerunning{MIPSBLEED\@: Uncovering Microarchitectural Timing Leaks}
\authorrunning{Ahmed Najeeb and Billy Bob Brumley}

\usepackage[T1]{fontenc}
\usepackage{graphicx}
\usepackage{xspace}

\newcommand{\PARAGRAPH}[2][.]{\medbreak\noindent\textbf{#2#1} }
\newcommand{\footurl}[1]{\footnote{\url{#1}}}

\title{MIPSBLEED\@: Uncovering Microarchitectural Timing Leaks in Pervasive Embedded Processors}

\usepackage{booktabs}
\usepackage{array}
\usepackage{makecell}
\begin{document}

\maketitle

\begin{abstract}
Despite their age, MIPS processors remain deeply embedded in routers, industrial controllers, and
IoT systems, yet their security against modern side-channel attacks has received little attention.
This paper exposes how Simultaneous Multithreading (SMT), a feature increasingly used to boost
performance in these environments, creates powerful cross-core timing channels on MIPS-based
platforms. We introduce MIPSBLEED, a systematic analysis and exploitation framework that uncovers
leakage in three shared microarchitectural components: the L1 data cache, L1 instruction cache, and
the execution engine. Through carefully crafted assembly-level probes and quantitative leakage
assessment, we demonstrate practical, high-resolution timing attacks that operate without requiring privileged
access. Our evaluation reveals significant information leakage across all
three channels and culminates in a single trace key recovery attack on a real elliptic curve
cryptographic toolkit. These results position MIPS as an overlooked yet critical target in the study
of microarchitectural security and underscore the urgent need for lightweight isolation mechanisms
in resource-constrained, SMT-enabled embedded systems.
 \keywords{
embedded security \and
applied cryptography \and
side-channel analysis \and
timing attacks \and
cache-timing attacks \and
microarchitectural attacks \and
MIPS}
\end{abstract}

\section{Introduction}\label{sec:intro}

Side-channel attacks have emerged as a powerful class of implementation exploits, capable of
extracting sensitive information by observing subtle variations in timing
\cite{DBLP:conf/crypto/Kocher96}, power \cite{DBLP:conf/crypto/KocherJJ99}, or various shared
resources \cite{bernstein2005cache}.
While much of the recent focus has been on
x86 \cite{DBLP:conf/sp/KocherHFGGHHLM019,DBLP:journals/cacm/LippSGPHHMKGYHS20,DBLP:conf/sp/LippKOSECG21,DBLP:conf/uss/WangPHSFK22,DBLP:conf/hpca/KimHB22,DBLP:conf/uss/PaccagnellaLF21},
and ARM \cite{DBLP:conf/uss/YuDJKF23,DBLP:conf/ccs/ZhangXZ16,DBLP:conf/uss/GreenLZIHE17,DBLP:conf/dac/BarenghiP18,DBLP:conf/acsac/Cronin0WC21} platforms,
the MIPS architecture remains a critical and underexamined target.
Despite its age, MIPS is far from obsolete:
it continues to power a vast array of embedded systems, networking hardware, and industrial controllers,
with an estimated 8.5 billion chips shipped as of 2024 \cite{mips}.
Its longevity, simplicity, and licensing model have made it a mainstay in
system-on-chip (SoC) designs for over two decades, and its architectural legacy directly influences
the design of modern and future open ISAs such as RISC-V \cite{DBLP:journals/cacm/HennessyP19}.
As such, analyzing microarchitectural side-channel leakage on MIPS is not only relevant but a critical
step toward understanding the broader landscape of hardware-level vulnerabilities.

This work presents a systematic exploration of timing-based side-channel
vulnerabilities in SMT-enabled MIPS processors. We focus on three shared
microarchitectural components: the L1 data cache, the L1 instruction cache, and the
execution engine. We demonstrate that all three can be exploited to leak
information across logical cores. These vulnerabilities arise from the
fundamental design of SMT \cite{Percival05}, which enables multiple hardware threads to share
execution resources without strong isolation guarantees.
While SMT is often
a desirable feature to improve throughput and resource utilization \cite{DBLP:conf/isca/TullsenEL95},
it also creates new
attack surfaces \cite{DBLP:conf/ches/AciicmezBG10,DBLP:conf/uss/GrasRBG18,DBLP:conf/sp/AldayaBHGT19}
that can be particularly applicable in embedded environments where
isolation boundaries are weaker due to the absence of trusted execution environments (TEE).

Our methodology builds on established low-level probing techniques, but adapts them specifically to
the MIPS architecture. We develop a suite of assembly-level tools that induce contention on shared
resources and measure the resulting latency variations. These probes are designed to be lightweight
and conceptually portable, enabling fine-grained characterization of leakage channels without
requiring privileged access or speculative execution (out of scope in this work). While similar
approaches have been used to study
x86 \cite{DBLP:journals/iacr/OsvikST05,DBLP:conf/sp/LiuYGHL15,DBLP:conf/uss/GrasRBG18,DBLP:conf/dac/KayaalpAPJ16,DBLP:conf/sp/GastJSSKFKG23}
and ARM \cite{DBLP:conf/uss/LippGSMM16},
our work systematically applies them to SMT-enabled MIPS
processors, where such leakage has not been characterized. This enables us to expose and quantify
side-channel vectors in this platform.

A key contribution of this work is the integration of quantitative leakage
assessment into the attack development process. While many side-channel studies
stop at demonstrating timing variation, we go further by evaluating whether the
observed signals are sufficient to support practical attacks. This step is
critical: without a quantitative measure of leakage, it is difficult to
distinguish between incidental noise and exploitable information flow. Our
analysis confirms that the timing variations induced by our probes are not only
measurable, but also carry sufficient information to enable fine-grained inference
of victim behavior.

To illustrate the practical implications of our findings, we apply our
techniques to real-world cryptographic tooling, recovering secret keys from
a widely deployed elliptic curve toolkit.
We mount our attack using only a single timing trace.
Yet this is just one
instantiation of our techniques, which are
applicable to a much wider range of scenarios.

\PARAGRAPH{Contributions}%
Our contributions are as follows:
\begin{itemize}
\item We present the first comprehensive side-channel analysis of SMT-enabled MIPS
processors, identifying exploitable leakage in the data cache, instruction
cache, and execution engine.
\item We integrate quantitative leakage assessment into the attack workflow,
demonstrating that the observed signals are sufficient to support practical
behavioral inference.
\item While general in nature, we demonstrate that our techniques
apply to real-world applications by carrying out an end-to-end key-recovery
attack on an unmodified version of a packaged open-source cryptographic application.
\end{itemize}

\PARAGRAPH{Outline}%
\autoref{sec:back} provides background on the MIPS microarchitecture, its SMT implementation,
and prior work on timing-based microarchitectural side channels, particularly in the SMT setting.
\autoref{sec:threatModel} defines our threat model and attacker capabilities.
In \autoref{sec:cache-attack} and \autoref{sec:ee}, we present our methodology for probing and measuring leakage originating from
the L1 data cache, L1 instruction cache, and execution engine,
including our leakage assessment framework, and we also present empirical results demonstrating the viability of each channel.
\autoref{sec:endAttack} discusses the application of our techniques to real-world cryptographic software, illustrating the relevance of our findings.
We conclude in \autoref{sec:conclusion} with a discussion of system security lessons, potential mitigations, and directions for future work.
\section{Background}\label{sec:back}

\subsection{Microarchitecture}\label{sec:backMicro}

This section examines major microarchitectural components of MIPS processors and their behavior under
MIPS's implementation of Simultaneous Multithreading (SMT). SMT leverages hardware threads to
achieve parallelism without duplicating all physical microarchitectural components typically present
in a dedicated physical processor. As a result, users experience the abstraction of two processors,
referred to as logical cores, within a single physical processor. These logical cores share most of
the underlying microarchitectural components, enabling efficient resource utilization while
maintaining parallel execution capabilities.

\begin{figure}[t]
\includegraphics[width=\linewidth]{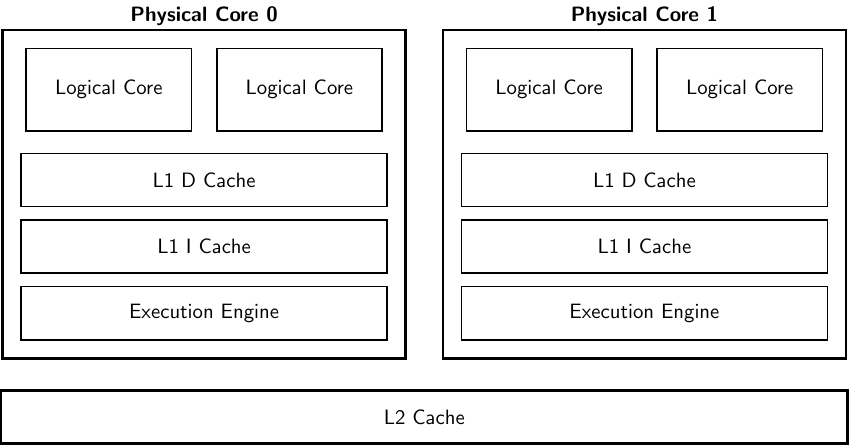}
\caption{High-level layout of the MIPS32 1004K processor.
Each physical core contains two logical cores sharing L1 data cache,
L1 instruction cache, and execution engine.}%
\label{fig:processor}

~\\

\includegraphics[width=\linewidth]{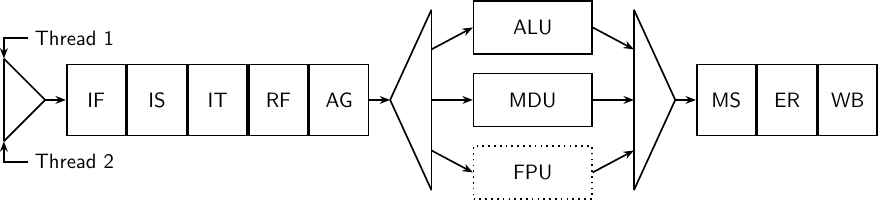}
\caption{MIPS32 1004K pipeline stages.
Instructions from both logical cores are interleaved
through the shared 8--9 stage execution engine.}%
\label{fig:pipeline}
\end{figure}

\autoref{fig:processor} shows a high-level overview of the layout of a MIPS32 processor. The
figure depicts two physical cores, each containing two logical cores. In this setting, the OS sees a
processor with four cores. Furthermore, as \autoref{fig:processor} shows, L1 instruction
cache and data cache are of equal size and are shared between the two logical cores alongside the
execution engine. The shared configuration is significant to our work as the shared nature of
these components result in side-channel attacks presented herein.

Caches help reduce the time it takes for the CPU to access data stored in memory. Instead of
fetching data from the slower main memory (RAM) every time, frequently used data is stored in the
cache, which is a smaller, faster memory located closer to the processor. This minimizes latency,
allowing the CPU to work more efficiently. However, caches are limited in size due to cost and
physical space constraints. Making a larger cache would increase latency and cost, reducing the
performance benefits, therefore as a compromise between the two factors multilevel caches are used.
The L1 cache in our instance of MIPS32 is 32KB, 4-way set associative with a line size of 32 bytes,
this gives us 1024 lines of L1 cache containing 256 sets. In MIPS32 the cache uses least recently
used (LRU) cache replacement policy to deal with cache conflicts \cite{MIPS-software-training}. Since both the logical cores share
the L1 cache and they both can write to it, this can lead to one process evicting the cache lines of the
other process running on the same physical core.

The execution engine is responsible for execution of instructions and is closely related to the
concept of pipelining in computer architecture. In our case the MIPS pipeline consists of 8--9 stages
as shown in \autoref{fig:pipeline}: (1) instruction fetch (IF), (2) instruction fetch second (IS),
(3) instruction fetch third (IT), (4) register file access (RF), (5) Address Generation (AG), (6) Execute/Memory Access (EX),
(7) Memory Access Second (MS), (8) Exception Resolution (ER), and (9) write back (WB)
\cite{MIPSsheet}.
The main goal of the pipeline is to increase instruction throughput by
executing multiple instructions simultaneously. Each logical core has its own register file, and
the pipeline fetches instructions from memory according to the program counter on each of them. For
the sake of processing performance fairness, this fetching is interleaved between the logical cores,
this means that instructions from both threads have to share the execution engine. The execution
engine of the MIPS processor contains two execution units: the ALU is responsible for all integer
arithmetic and logical operations, and the MDU (Multiply/Divide Unit) is responsible for multiplication and
division operations. In some variants like the 1004KF, the execution engine includes a
floating-point unit (FPU), but our variant does not have this feature \cite{MIPSsheet}.

\PARAGRAPH{Comparison with ARM and x86}%
As illustrated in \autoref{tab:arch_comparison}, the MIPS32 architecture adopts a distinctive strategy for supporting
high-fidelity timing and cache management compared to ARM and x86 platforms. To measure the
clock cycles of each operation we use the RDHWR (Read Hardware Register) instruction. The
instruction moves the contents of a hardware register to a general purpose register. Specifically,
in this context, the High-Resolution Cycle Counter (CC), located within coprocessor 0, is accessed
to obtain precise cycle counts. Unlike the ARM architecture, which necessitates privileged mode to
access the CC, or the x86 architecture, where the CC is accessible from user space via a dedicated
built-in RDTSC instruction, MIPS32 takes a different approach. MIPS allows access to CC from user
space by default, however this behavior can be restricted by a privileged user limiting user-space
access to the CC\@. Furthermore there is no instruction in MIPS to flush the cache similar to x86 nullifying
attacks such as Flush+Reload \cite{DBLP:conf/uss/YaromF14}. Another aspect of MIPS which is similar to x86
is the cache replacement policy, MIPS32 uses a Least Recently Used (LRU) cache replacement policy
similar to x86, while ARM uses a pseudo-random replacement policy \cite{DBLP:conf/uss/LippGSMM16}.

\begin{table}[h!]
\centering
\caption{Comparison of user-space timing, cache flush, and
cache replacement features across x86, MIPS32, and ARM architectures.}%
\label{tab:arch_comparison}
\begin{tabular}{@{} lccc @{}}
    \toprule
    \textbf{Feature} & \textbf{x86} & \textbf{MIPS} & \textbf{ARM} \\
    \midrule
    \makecell[l]{User-Space High-\\Fidelity Timing} & \texttt{rdtsc} & \texttt{rdhwr} & N/A \\
    Cache Flush & \texttt{clflush} & N/A & N/A \\
    Cache Replacement Policy & LRU & LRU & Pseudo Random \\
    \bottomrule
\end{tabular}
\end{table}

\subsection{SMT\@: Timing Attacks}\label{sec:backSMT}

Simultaneous Multithreading (SMT) technology, introduced to enhance the efficiency of
microprocessors, has also exposed systems to timing attacks, a key part of side-channel \cite{Percival05,DBLP:journals/iacr/AciicmezSK06,DBLP:journals/iacr/GeYCH16}
analysis. By allowing multiple threads to share hardware resources like caches and execution units,
SMT can lead to resource contention, causing measurable timing variations. These variations can
unintentionally reveal sensitive information, serving as both covert and side channels for
attackers.

\PARAGRAPH{Data Cache}%
In his groundbreaking work, Percival~\cite{Percival05} presented a novel cache-timing attack against RSA's Sliding Window
Exponentiation (SWE) implemented in OpenSSL 0.9.7c. He demonstrated that since two threads share the same
L1 data cache and Intel processors were using LRU cache policy, one thread can evict cache lines of the other
thread. In the attack instance a spy would occupy the entire cache and time each cache set access, a delay
would indicate eviction, leaking information about the colocated thread.
The behavior leads to a side channel that---when exploited in the context of OpenSSL SWE implementation---can
lead to correct identification of precomputed multipliers resulting in RSA private key recovery. As a
countermeasure to this attack OpenSSL included a ``constant time'' implementation of the SWE algorithm.

At the same time, Osvik, Shamir, and Tromer~\cite{DBLP:journals/iacr/OsvikST05} demonstrated the use of their Prime+Probe
technique to carry out an attack on an AES implementation in OpenSSL\null. In Prime+Probe the attacker first primes the cache by
filling it with their own data, then they wait for the victim to execute and finally they probe the cache
by accessing the same cache lines. The difference in access time indicates if the victim evicted
the attacker's data from the cache, in this case a delay indicates eviction. Using this technique, the authors
were successfully able to recover the AES key from OpenSSL 0.9.8's implementation.

Building on the Prime+Probe primitive and work by Gullasch, Bangerter, and Krenn~\cite{DBLP:conf/sp/GullaschBK11},
Yarom and Falkner~\cite{DBLP:conf/uss/YaromF14} introduced Flush+Reload as a high-resolution
cache side-channel attack specifically targeting the last-level (L3) cache, demonstrating the evolution
of cache timing attacks. Unlike Prime+Probe, which infers cache activity by priming sets and probing for evictions,
Flush+\-Re\-load leverages shared memory pages such as those arising from shared libraries or memory deduplication
to monitor specific cache lines with much finer granularity. By flushing a targeted cache line using the clflush
instruction and subsequently reloading it while measuring access time, the attacker can precisely
determine whether the victim accessed the line in the interim. If the reload is fast, the line is
present in the cache, implying victim access; otherwise, it must be fetched from main memory,
indicating no access. Using this technique, the authors were able to recover private encryption keys from
GnuPG 1.4.13's RSA implementation. This attack is not possible on MIPS as it does not have the clflush instruction or
similar functionality accessible from user space.

\PARAGRAPH{Instruction Cache}%
Ac\i{}i\c{c}mez~\cite{DBLP:journals/iacr/Aciicmez07} in his pioneering work showcased the first use
of the Instruction Cache (Icache) timing side-channel attack. The attack is based on the fact that a spy
is able to stop the execution of the victim process just before the execution of the ``spied-on''
part of cipher code and fill up the Icache set with the dummy instructions. The spy then resumes the victim
process and times the execution of the cipher code which evicts some of the dummy instructions. When the spy executes
again it times the access to dummy instructions and if there is a delay it can infer the evicted cache sets. By increasing the
number of interruption to the victim process and adding dummy instructions for different sets, the spy can determine
which sets the victim is using. Leveraging this information the authors determine that the attack on OpenSSL's RSA
SWE implementation is feasible.

Building on \cite{DBLP:journals/iacr/Aciicmez07}, Ac\i{}i\c{c}mez, Brumley, and Grabher~\cite{DBLP:conf/ches/AciicmezBG10} demonstrate that
timing side-channel attacks on Icache are not only feasible but also practical. The authors adapt the eviction
strategy from Percival~\cite{Percival05} to the Icache setting. The strategy is to pollute the Icache with the Spy's instructions and time the
latency of code execution on Intel's Atom processor.
Applying an existing methodology to process cache-timing data \cite{DBLP:conf/asiacrypt/BrumleyH09},
the authors then pass the timing data into a Vector Quantization (VQ) model which
classifies the data based on a vector codebook obtained during the profiling stage of the attack. The output of the VQ model is then
fed into a hidden Markov Model (HMM) after which the Viterbi algorithm is used to predict the most likely state sequence. Finally, a lattice
attack is carried out to recover the DSA key, demonstrating that the attack is able to leak critical state from OpenSSL 0.9.8l's implementation.

\PARAGRAPH{Execution Engine}%
Besides the data and instruction cache, other shared microarchitectural components such as execution
engines are also vulnerable to timing attacks.
Aldaya et al.~\cite{DBLP:conf/sp/AldayaBHGT19} carry
out their PortSmash attack on Intel Skylake and Kaby Lake chips. The Intel microarchitecture uses ports to
schedule instructions for the execution engine. Essentially, ports are a channel to stacks of
execution units in the same way that network ports can be channels to different daemons. The authors
create a high-resolution timing side-channel due to port
contention, enabled by SMT\@. They demonstrate the recovery of a P-384 elliptic curve private key from an OpenSSL
1.1.0h powered TLS server using a small number of repeated TLS handshake attempts.
In a similar vein, Bhattacharyya et al.~\cite{DBLP:journals/corr/abs-1903-01843} exploit port contention to carry out a speculative code-reuse attack leaking
sshd private keys in OpenSSL libcrypto library (version 1.1.1b).

Another example of instruction-level timing leakage arises from the variable latency of division instructions.
Bernstein et al.~\cite{DBLP:journals/iacr/BernsteinBBCCKKPRT24}
utilize this fact to demonstrate that typically a division with a constant is optimized to multiplication but in some cases such as asking
the compiler to optimize for code size, the optimization is disabled, creating a timing side-channel.
In their KyberSlash attack, the authors are able to successfully leak information
about the secret key and ciphertext in the Kyber Post-Quantum Key Encapsulation Mechanism reference code.

\subsection{Leakage Assessment}\label{sec:nicv}

Several statistical techniques are commonly used in side-channel analysis (SCA)
to evaluate information leakage, including Pearson's correlation coefficient \cite{DBLP:conf/fc/CoronKN00},
Welch's T-test, Test Vector Leakage Assessment (TVLA)  \cite{Welch}, and Normalized
Inter-Class Variance (NICV) \cite{NICV-Paper}. These methods facilitate the detection of leakage
in collected traces for SCA\@.

Pearson's correlation coefficient quantifies the linear relationship between two
random variables and is widely employed in evaluating information leakage \cite{DBLP:conf/fc/CoronKN00}.
It is
particularly useful for identifying Points of Interest (POIs) within
side-channel traces, as seen in template attacks \cite{TemplateAttacks}.

Welch's T-test is a statistical method used to assess if two sample
sets originate from populations with equivalent means.
Goodwill et al.~\cite{Welch}
introduced Test Vector Leakage Assessment (TVLA), which applies the T-test to
evaluate information leakage by comparing trace sets generated using fixed
versus random cryptographic keys and input data.

Bhasin et al.~\cite{NICV-Paper} proposed NICV as a method for leakage
assessment. It is based on the Analysis of Variance (ANOVA) F-test, a
statistical approach used to determine whether multiple sample sets originate
from populations with comparable variances. We use NICV as our metric to assess leakage.

\section{Threat Model}\label{sec:threatModel}

At a high level, our threat model is similar to \cite{DBLP:conf/ches/AciicmezBG10,DBLP:journals/iacr/YaromGH16,DBLP:conf/uss/GrasRBG18}.
We assume an attacker capable of executing unprivileged code on the victim's
system, with SMT enabled \cite{DBLP:journals/corr/abs-1903-01843}.
For all proposed attacks to be effective, the attacker
and victim must be co-located on the same physical
processor core \cite{DBLP:journals/iacr/GeYCH16},
thereby sharing the L1 data cache, L1 instruction cache, and
execution engine.
Under this co-residency condition, the attacker can launch any
of the three side-channel attacks described in this work, targeting scenarios
where the victim is processing sensitive information such as cryptographic keys
or passwords.

All of our attacks leverage latency measurements to detect information leakage.
Specifically, the attacker executes a sequence of instructions designed to
saturate a shared resource, such as the cache or execution engine, while
continuously measuring execution time. Any activity from the victim process
introduces contention, increasing the observed latency for the attacker. By
analyzing these fluctuations, the attacker can infer which specific regions of
memory or execution units the victim is utilizing, leading to sensitive
information leakage.
\section{Cache Side-Channel}\label{sec:cache-attack}

The L1 cache geometry of the MIPS processor used in our setup is as follows:
32 byte cache lines, 32KB total size, 4-way set associative, and 1024 lines divided
into 256 associative sets. Each address is split into three parts; offset, index, and tag.
The offset is 5 bits used to indicate one of the 32 bytes in each line. The index consists
of 8 bits used to indicate one of the 256 sets, and the tag contains the remaining 19 bits
to distinguish between different memory regions.

\subsection{Data Cache Side-Channel}\label{sec:dcache}

\begin{figure}[!t]
\includegraphics[width=\linewidth]{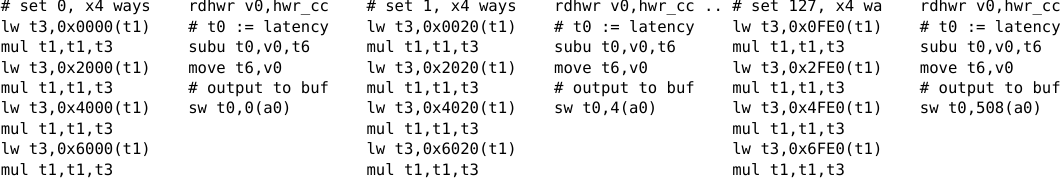}
\caption{L1 data cache probe implementation excerpt (MIPS assembly).
Three iterations shown for cache sets 0, 1, and 127, each
accessing all four ways with a dependency chain to prevent out-of-order execution.}%
\label{fig:dcacheProbe}
\end{figure}

\begin{figure}[!b]
\includegraphics[width=\linewidth]{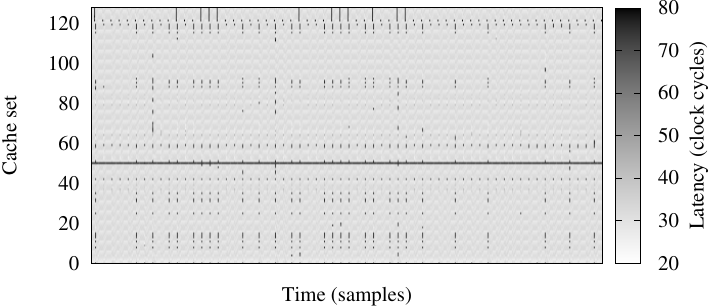}
\caption{L1 data cache probe output with the victim accessing set 50.
The \(y\)-axis denotes cache set index (128 sets monitored) and
the \(x\)-axis denotes successive probe iterations.
The horizontal dark band at set 50 confirms measurable contention.}%
\label{fig:dcacheProbee}
\end{figure}

\begin{figure}[!t]
\includegraphics[width=\linewidth]{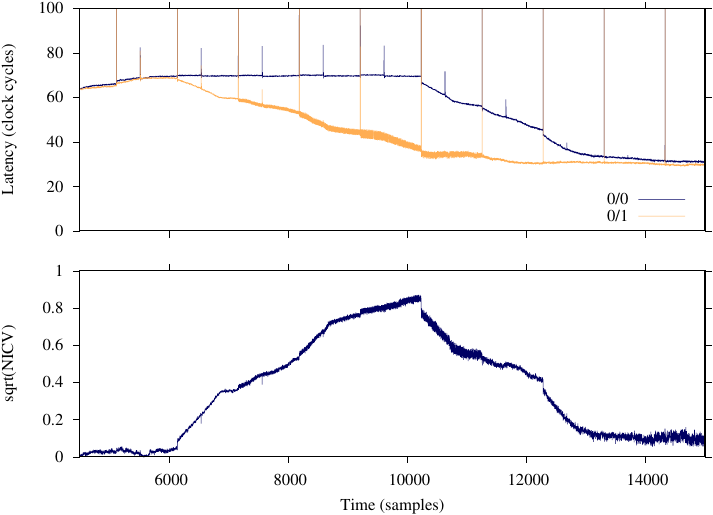}
\caption{Top: averaged L1 data cache traces across two classes
(sustained contention vs.\ contention then idle).
Vertical lines correspond to context switches.
Bottom: corresponding NICV values; high NICV confirms significant
information leakage through the data cache channel.}%
\label{fig:dcacheNICV}
\end{figure}

The attacker issues a large number of load accesses to the L1 data cache and
measures the time required to carry out each access. During the attack, if the
victim also accesses the data cache, the attacker's cache access time will
increase due to contention for the same cache lines \cite{DBLP:journals/iacr/OsvikST05}. This increase in latency
reveals the cache lines accessed by the victim.

We implemented our attack in assembly, and \autoref{fig:dcacheProbe} shows an excerpt.
As discussed in \autoref{sec:backMicro}, the MIPS processor in our setup has a 4-way
set-associative cache comprising 256 sets with a line size of 32 bytes.
Consequently, our spy executes 256 iterations, accessing four cache lines per
iteration. Each cache line within a set is offset by 8KB, leading to the
selection of the base addresses shown.

The set-0-aligned input buffer (\verb+t1+) is initialized with all ones, effectively causing
the mul operation to function as a no-operation (NOP) instruction. This approach
enforces a dependency chain in cache accesses, preventing out-of-order
execution (inspired by \cite{DBLP:conf/cosade/BrumleyT11}).
Additionally, we use different registers to avoid register
contention, ensuring that it does not affect our measurements. Finally, once per
set iteration, we record the value of the cycle counter and compute the
difference from the previous iteration. This difference corresponds to the
measured access latency, which is then stored in an output array for analysis.

\autoref{fig:dcacheProbe} illustrates three such iterations, for sets 0, 1, and 127.
In the excerpt, \verb+t1+ is a 13-bit aligned array, and all values contain the integer 1 as a 4-byte integer.
The cache set 0 column (left) is almost identical to the set 1 column (center),
the only differences being an extra 32 bytes in each fixed load offset (lines are 32 bytes) to select the cache set,
and the fixed store offset increase to account for the persisted latency measurements.
Cache set 127 (and all other cache sets) follows this pattern.
In the full code, we have control flow logic to gather successive measurements (i.e., loop) following the final cache set.

To validate the effectiveness of our probe code and confirm the presence of cache contention, we
perform an experiment in which a victim process repeatedly accesses a specific cache set (here, set 50) while the
attacker concurrently executes the probe code to monitor access latencies.
In this experiment, we only measure the first half of the data cache (i.e., 128 sets) to increase temporal resolution.
\autoref{fig:dcacheProbee} illustrates the result of this experiment.
The \(y\)-axis denotes the cache set index, and the \(x\)-axis
represents time over successive probe iterations. As shown, a clear increase in access latency is
consistently observed at cache set 50, evidenced by the horizontal streak of darker pixels centered on
that set. This elevated latency pattern confirms that the victim's accesses interfere with the
attacker's probe, validating the probe's ability to detect contention at a fine granularity. This
result demonstrates that our probing code can reliably identify cache sets targeted by a concurrent
victim process, meaning data cache is an effective side-channel in MIPS\@.

\subsection{Instruction Cache Side-Channel}\label{sec:icache}

\begin{figure}[!b]
\includegraphics[width=\linewidth]{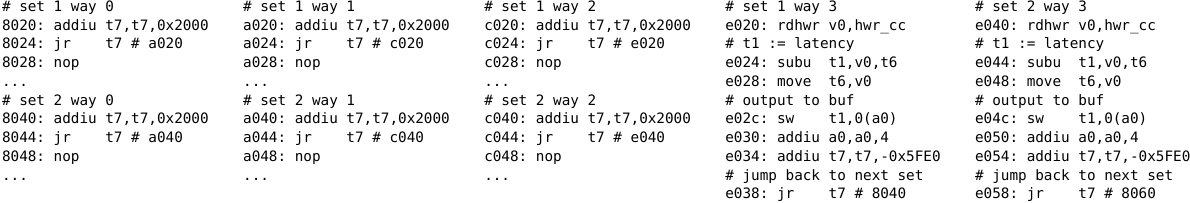}
\caption{L1 instruction cache probe implementation excerpt (MIPS assembly).
Two iterations shown for cache sets 1 and 2, each jumping through all
four ways at 8KB offsets before recording the latency measurement.}%
\label{fig:IcacheProbe}
\end{figure}

\begin{figure}[!t]
\includegraphics[width=\linewidth]{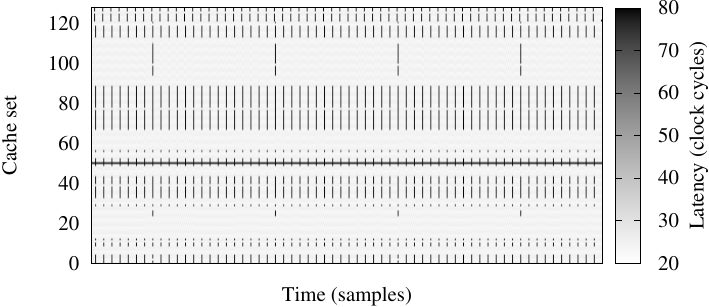}
\caption{L1 instruction cache probe output with the victim executing instructions aligned to set 50.
The \(y\)-axis denotes cache set index (128 sets monitored) and the \(x\)-axis denotes successive probe iterations.
The horizontal dark band at set 50 confirms measurable contention.}%
\label{fig:IcacheProbee}

\bigskip

\includegraphics[width=\linewidth]{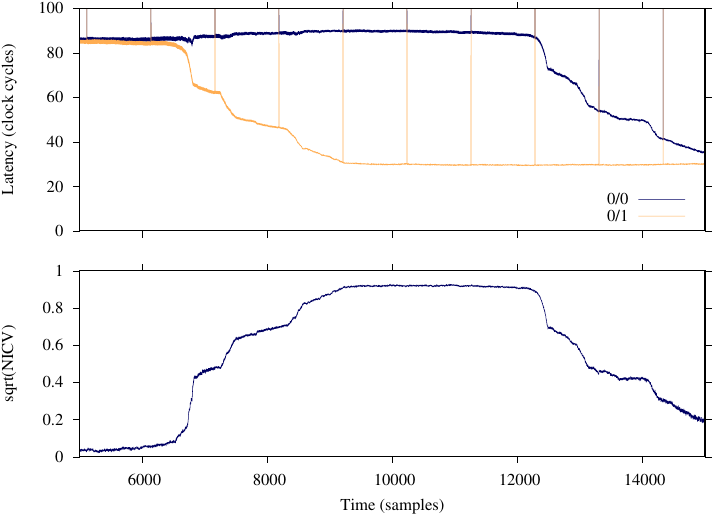}
\caption{Top: averaged L1 instruction cache traces across two classes
(sustained contention vs.\ contention then idle).
Vertical lines correspond to context switches.
Bottom: corresponding NICV values; high NICV confirms significant
information leakage through the instruction cache channel.}%
\label{fig:IcacheNICV}
\end{figure}

The attacker issues a large number of instructions, flooding the L1 instruction
cache and measures the time required to execute said instructions through the cache.
During the attack, if the victim also accesses the instruction cache, the
attacker's cache access time will increase due to contention for the same cache
lines \cite{DBLP:conf/ches/AciicmezBG10}. This increase in latency reveals the cache lines accessed by the victim.

\autoref{fig:IcacheProbe} shows an excerpt of our assembly implementation.
It shows two iterations, one for cache set 1 and then follows with another for set 2.
This logic repeats for each of the 256 cache sets
(see \autoref{sec:backMicro}), taking a timing measurement for each set. The code is
structured in a similar way to \cite{DBLP:conf/ches/AciicmezBG10}. First we
align our code at a 13-bit boundary (here \verb+8000+, and note \verb+t7+ is initially set to the corresponding virtual address),
making it easier to fill up the entire
4-way cache in contiguous 32-byte regions of code. We then access the
instruction cache in a certain pattern that evicts each cache set.

To illustrate, we step through the control flow in \autoref{fig:IcacheProbe}.
Assume the cache set 0 measurement completed, and the code now jumps to \verb+8020+,
i.e., cache set 1 and way 0. The code adds 8KB to the jump target and jumps through the register (\verb+t7+),
arriving at \verb+a020+, i.e., cache set 1 and way 1.
It then adds another 8KB and jumps to arrive at \verb+c020+, i.e., cache set 1 and way 2.
It then adds another 8KB and jumps to arrive at \verb+e020+, i.e., cache set 1 and way 3.
Since the spy has now exhausted all four ways, it stores the latency measurement and continues with cache set 2,
subtracting off the added 24KB and then increasing by an extra 32 bytes to increment the cache set from 1 to 2.
The code then jumps to \verb+8040+, i.e., cache set 2 and way 0.
The code adds 8KB to the jump target and jumps through the register (\verb+t7+),
arriving at \verb+a040+, i.e., cache set 2 and way 1.
It then adds another 8KB and jumps to arrive at \verb+c040+, i.e., cache set 2 and way 2.
It then adds another 8KB and jumps to arrive at \verb+e040+, i.e., cache set 2 and way 3.
Since the spy has now exhausted all four ways, it stores the latency measurement and continues with cache set 3 (\verb+8060+).
In the full code, we have control flow logic to gather successive measurements (i.e., loop) following the final cache set.

To validate our Icache probe code, we run an experiment in which a victim process executes
instructions specifically aligned to conflict with set 50 of the L1 instruction cache, while the
attacker uses the full \autoref{fig:IcacheProbe} code to measure set-wise latency.
In this experiment, we only measure the first half of the instruction cache (i.e., 128 sets) to increase temporal resolution.
This setup allows us to test whether our
probing technique can detect contention caused by concurrent instruction fetches. As shown in
\autoref{fig:IcacheProbee}, a clear and persistent increase in access latency is observed at set 50
across time, indicating that the victim's activity in the instruction cache introduces measurable
contention. This spike confirms that our probe is both accurately targeting individual cache sets
and sensitive enough to capture interference caused by external instruction execution. Thus, the
experiment not only demonstrates that the instruction cache can be a viable side-channel on our MIPS
platform, but also confirms that the probe code correctly reveals cache sets accessed by the victim.

\subsection{Results}\label{sec:cacheResults}

The results of the data and instruction cache attacks are shown in
\autoref{fig:dcacheNICV} and \autoref{fig:IcacheNICV}, respectively. For both
cases, NICV is measured using the averaged latency for each trace across two
classes: the first class involves sustained cache contention, while the second
class involves cache contention followed by no contention \cite{DBLP:conf/uss/AldayaB22}. In the second class,
the victim is not executing any code affecting the probed cache lines during the latter half of the trace,
this behavior is illustrated in the top halves of \autoref{fig:dcacheNICV} and \autoref{fig:IcacheNICV}.
The vertical lines across both figures are a result of context switches during the probe.
The bottom halves of both figures show the corresponding NICV values for probed data and instruction caches, with high NICV
indicating significant information leakage through the channel.
\section{Execution Engine Side-Channel}\label{sec:ee}

\begin{figure}[!t]
\includegraphics[width=\linewidth]{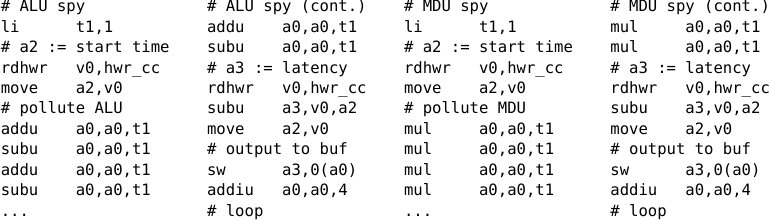}
\caption{Execution engine probe implementation (MIPS assembly).
Left: ALU contention via repeated add/sub pairs.
Right: MDU contention via repeated mul instructions.
Both record per-iteration cycle counts.}%
\label{fig:eeProbe}

\bigskip

\includegraphics[width=\linewidth]{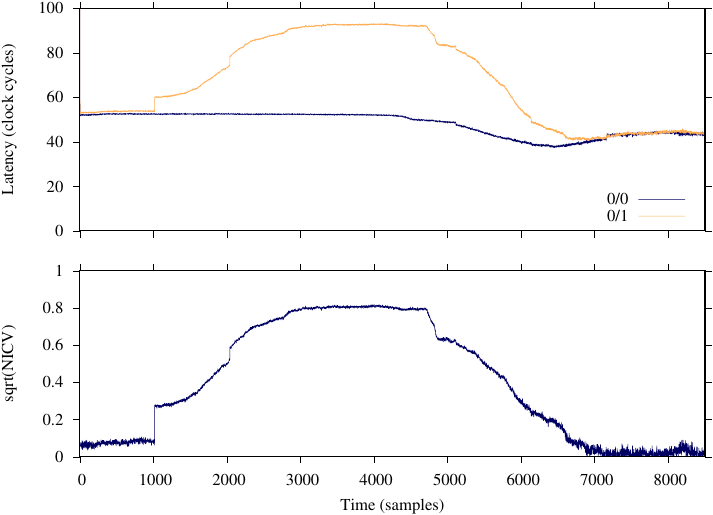}
\caption{Top: averaged execution engine traces across two classes
(no contention vs.\ ALU contention in the second half).
Bottom: corresponding NICV values; high NICV confirms significant
information leakage through the execution engine channel.}%
\label{fig:eeNICV}
\end{figure}

The attacker issues a large number of instructions saturating the execution
engine and measures the time required to retire these instructions.
During the attack, if the victim also issues a similar instruction, the
attacker's execution time will increase due to contention for the same execution
unit. This increase in latency reveals the execution units accessed by the victim,
an indication of the types and order of instructions being issued by the victim.

The assembly code shown in \autoref{fig:eeProbe} (left) is an excerpt of our implementation. In the
attack we repeat a pair of add and sub instructions with a latency of 1, 25 times, saturating the execution
engine. We take the timing before and after these repeated instructions to calculate latency and store
it into an output array. It is important to note that for this attack any instruction can be used
to measure latency depending upon the victim's execution pattern.
The \autoref{fig:eeProbe} (right) code is similar, yet targets the MDU with mul instructions rather than the ALU\null.

\subsection{Results}\label{sec:eeResults}

The results of the attack are shown in \autoref{fig:eeNICV}. For measuring NICV we utilize the
averaged latency for each trace across two classes. The first class is addition throughout with no
contention while the second class is no contention then ALU contention in the second half \cite{DBLP:conf/uss/AldayaB22}. This result can be
seen in \autoref{fig:eeNICV} top half. There are no visible lines showing context switches as execution
engines are agnostic to them.

The bottom half of \autoref{fig:eeNICV} shows the NICV for the execution engine probe. It can be seen that
NICV is very high, demonstrating strong leakage across the channel.
\section{End-to-End Attack}\label{sec:endAttack}

In this section, we demonstrate real-world relevance of our MIPS-based side-channel attacks
by carrying out an end-to-end key recovery attack on an unmodified open-source, packaged version
of a cryptographic tool.

\subsection{The SECCURE Toolset}\label{sec:seccure}
The Secure Elliptic Curve Crypto Utility for Reliable Encryption
(SECCURE) is a lightweight elliptic curve cryptography (ECC) toolkit that
provides a command-line interface for a range of cryptographic
operations. Its functionality includes key generation (seccure-key), public-key
encryption and decryption (seccure-en\-crypt and seccure-decrypt), digital
signature generation and verification (seccure-sign and seccure-verify), and
support for Diffie-Hellman key exchange (seccure-dh). SECCURE supports
several standardized elliptic curves and is specifically optimized for
performance in resource-constrained environments, such as embedded systems.
Originally released in July 2006---prior to widespread awareness of side-channel
vulnerabilities---SECCURE remains actively available, with version 0.5 released
in August 2014, employed unaltered in our experiments. SECCURE is
available\footurl{http://point-at-infinity.org/seccure/}
as a package in mainstream Linux distributions, including Debian and Ubuntu. Internally,
SECCURE implements custom elliptic curve arithmetic and leverages libgcrypt via dynamic
linking to perform multiprecision arithmetic operations.

\subsection{Key-Recovery Attack}\label{sec:keyRecovery}

One security-critical operation with ECC is scalar multiplication, \(kP\),
which adds the point \(P\) to itself \(k\) times (an integer).
There are many implementation strategies for scalar multiplication,
that are typically based on repeated doublings and conditional additions.
Scalar multiplication is analogous to modular exponentiation in a multiplicative setting,
where doublings are squarings and additions are multiplications.

\begin{figure}[t]
\centering
\includegraphics[width=0.7\linewidth]{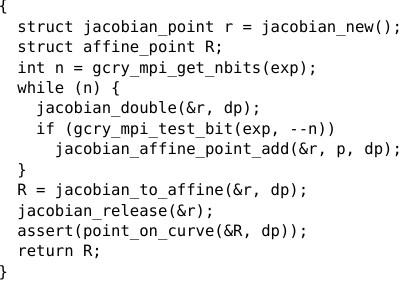}
\caption{Vulnerable code segment in SECCURE's \texttt{pointmul} function.
The conditional branch on each scalar bit introduces a
data-dependent point addition, creating exploitable timing variation.}%
\label{fig:seccureCode}

~\\

\includegraphics[width=\linewidth]{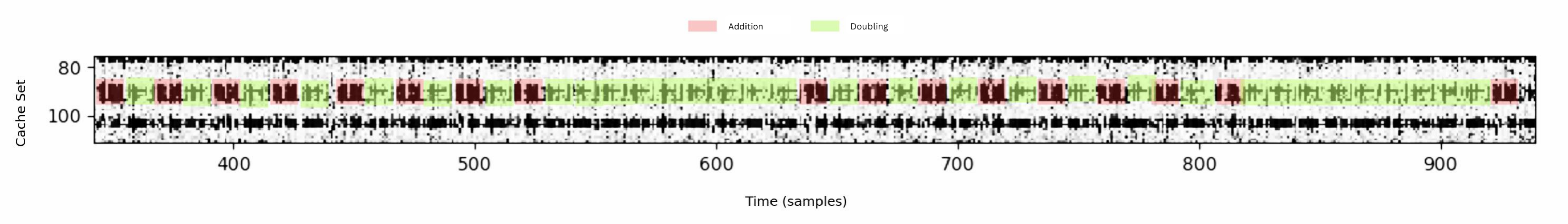}
\caption{Partial instruction cache trace during SECCURE scalar multiplication on P-256.
Darker regions indicate higher latency.
Red and green highlights mark point addition and point doubling operations,
respectively, directly revealing scalar bits.}%
\label{fig:attackTrace}
\end{figure}

\autoref{fig:seccureCode} illustrates the vulnerable code segment within
SECCURE's implementation of scalar multiplication, specifically within the
pointmul function. The vulnerability arises from a conditional branch that
performs an elliptic curve point addition only when a corresponding bit in the
scalar is set. This data-dependent control flow introduces timing variation that
can be exploited via side-channel analysis.

The objective of the attack is to recover the sequence of elliptic curve point
doubling and addition operations executed during scalar multiplication. Since
this sequence is directly correlated with the scalar value, its recovery enables
full reconstruction of the private key. To carry out the attack, we execute the seccure-key
utility using the P-256 elliptic curve secp256r1, concurrently with a
cache probing routine on the same processor core. Our attack monitors the instruction cache activity
to infer the program's execution path. The resulting, partial instruction cache trace is shown in \autoref{fig:attackTrace}.

The trace reveals the direct leakage of the scalar multiplication
routine. In the trace, which plots cache set access latency over time, darker shades indicate higher
memory access latencies. The scalar multiplication algorithm performs a point doubling for every bit
of the scalar and a conditional point addition for every 1 bit shown in \autoref{fig:seccureCode}. This creates two distinct execution
paths, which are distinguishable in the cache trace. The recurring pattern of alternating dark and
light regions corresponds to the code path for a point addition followed by a point
doubling (highlighted in red and green, respectively). A continuous sequence of lighter regions (highlighted in green) would
represent the path for a point doubling only.

The partial trace therefore clearly shows a sequence of eight consecutive ``doubling and addition''
operations. Since each such operation corresponds to a 1 bit in the private key, we can infer that
the initial bits of the scalar are 11111111. This result is consistent with our experimental setup;
to create a clear and unambiguous signal for this demonstration, we modified the SECCURE
hash\_to\_exponent function to deterministically generate a key with a highly regular bit pattern
\(1111111100000000\ldots\) instead of a random one. This specific, predictable key was chosen to
highlight the side-channel leakage, and this partial trace confirms that the initial eight bits are
recovered successfully alongside the pattern being repeated, validating the attack. We emphasize
that our end-to-end techniques are fully capable of operating on normal, randomly generated keys;
however, we selected this predictable key pattern strictly for demonstration purposes to make the visual
evidence of the leakage unambiguous. Furthermore, the true magnitude and resolution of the
information leakage extend far beyond this simplified visual demonstration, as quantified by the
high NICV scores obtained in \autoref{sec:cache-attack}.

\section{Conclusion}\label{sec:conclusion}

In this work, we presented a systematic analysis of microarchitectural timing-based side-channel
vulnerabilities in SMT-enabled MIPS processors, demonstrating that shared
microarchitectural components---such as the L1 data cache, L1 instruction cache, and the
execution engine---can be exploited to leak sensitive information across
logical cores. Our methodology combines low-level probing with quantitative
leakage assessment, offering a framework for identifying and validating
side-channel vectors in embedded systems.

These findings have immediate relevance for MIPS-based platforms, which remain
widely deployed in long-lived, high-assurance environments. More broadly, our results
offer architectural insights applicable to emerging ISAs such as RISC-V, where
similar trade-offs between performance and isolation are being actively
explored. Our work provides both a warning and a toolkit for assessing and
addressing these challenges on emerging architectures.

Looking forward, we see several promising directions for future work. First, our
leakage primitives can serve as building blocks for more advanced attacks,
including speculative and transient execution techniques that exploit
timing attack vectors to bypass privilege boundaries. Second, extending our
methodology to other architectures and integrating it into automated testing
pipelines could support continuous leakage assessment during hardware and
firmware development. Finally, exploring lightweight mitigations in the embedded space
specific to SMT, both architectural and software-based.

To conclude, our work highlights the importance of treating microarchitectural
leakage as a first-class security concern in embedded systems, and provides a
foundation for side-channel security assessment in this setting.

\begin{credits}
\subsubsection{\discintname}
The authors have no competing interests to declare that are
relevant to the content of this article.
\end{credits}

\bibliographystyle{splncs04}
\bibliography{manuscript_rw,manuscript_ro}

@InProceedings{DBLP:conf/crypto/Kocher96,
  author        = {Paul C. Kocher},
  editor        = {Neal Koblitz},
  title         = {Timing Attacks on Implementations of {Diffie-Hellman}, {RSA}, {DSS}, and Other Systems},
  altbooktitle  = {{CRYPTO}},
  booktitle     = {Advances in Cryptology - {CRYPTO} '96, 16th Annual International Cryptology Conference, Santa Barbara, California, USA, August 18-22, 1996, Proceedings},
  series        = {Lecture Notes in Computer Science},
  volume        = {1109},
  pages         = {104--113},
  publisher     = {Springer},
  year          = {1996},
  url           = {https://doi.org/10.1007/3-540-68697-5_9},
  ALTdoi        = {10.1007/3-540-68697-5_9}
}

@InProceedings{DBLP:conf/crypto/KocherJJ99,
  author        = {Paul C. Kocher and Joshua Jaffe and Benjamin Jun},
  editor        = {Michael J. Wiener},
  title         = {Differential Power Analysis},
  altbooktitle  = {{CRYPTO}},
  booktitle     = {Advances in Cryptology - {CRYPTO} '99, 19th Annual International Cryptology Conference, Santa Barbara, California, USA, August 15-19, 1999, Proceedings},
  series        = {Lecture Notes in Computer Science},
  volume        = {1666},
  pages         = {388--397},
  publisher     = {Springer},
  year          = {1999},
  url           = {https://doi.org/10.1007/3-540-48405-1_25},
  ALTdoi        = {10.1007/3-540-48405-1_25}
}

@InProceedings{DBLP:conf/ccs/ZhangXZ16,
  author        = {Xiaokuan Zhang and Yuan Xiao and Yinqian Zhang},
  editor        = {Edgar R. Weippl and Stefan Katzenbeisser and Christopher Kruegel and Andrew C. Myers and Shai Halevi},
  title         = {Return-Oriented Flush-Reload Side Channels on {ARM} and Their Implications for {Android} Devices},
  altbooktitle  = {{ACM} {CCS}},
  booktitle     = {Proceedings of the 2016 {ACM} {SIGSAC} Conference on Computer and Communications Security, Vienna, Austria, October 24-28, 2016},
  pages         = {858--870},
  publisher     = {{ACM}},
  year          = {2016},
  url           = {https://doi.org/10.1145/2976749.2978360},
  ALTdoi        = {10.1145/2976749.2978360},
  bibsource     = {dblp computer science bibliography, https://dblp.org}
}

@InProceedings{DBLP:conf/asiacrypt/BrumleyH09,
  author        = {Billy Bob Brumley and Risto M. Hakala},
  editor        = {Mitsuru Matsui},
  title         = {Cache-Timing Template Attacks},
  altbooktitle  = {{ASIACRYPT}},
  booktitle     = {Advances in Cryptology - {ASIACRYPT} 2009, 15th International Conference on the Theory and Application of Cryptology and Information Security, Tokyo, Japan, December 6-10, 2009. Proceedings},
  series        = {Lecture Notes in Computer Science},
  volume        = {5912},
  pages         = {667--684},
  publisher     = {Springer},
  year          = {2009},
  url           = {https://doi.org/10.1007/978-3-642-10366-7_39},
  ALTdoi        = {10.1007/978-3-642-10366-7_39}
}

@InProceedings{Percival05,
  author        = {Colin Percival},
  title         = {Cache Missing for Fun and Profit},
  altbooktitle  = {{BSDCan}},
  booktitle     = {BSDCan 2005, Ottawa, Canada, May 13-14, 2005, Proceedings},
  year          = {2005},
  url           = {http://www.daemonology.net/papers/cachemissing.pdf}
}

@Article{DBLP:journals/cacm/HennessyP19,
  author        = {John L. Hennessy and David A. Patterson},
  title         = {A new golden age for computer architecture},
  journal       = {Commun. {ACM}},
  volume        = {62},
  number        = {2},
  pages         = {48--60},
  year          = {2019},
  url           = {https://doi.org/10.1145/3282307},
  ALTdoi        = {10.1145/3282307},
  bibsource     = {dblp computer science bibliography, https://dblp.org}
}

@InProceedings{DBLP:conf/cosade/BrumleyT11,
  author        = {Billy Bob Brumley and Nicola Tuveri},
  title         = {Cache-timing attacks and shared contexts},
  pages         = {233--242},
  altbooktitle  = {{COSADE}},
  booktitle     = {Constructive Side-Channel Analysis and Secure Design - 2nd International Workshop, COSADE 2011, Darmstadt, Germany, February 24-25, 2011. Proceedings},
  year          = {2011},
  oldurl        = {https://tutcris.tut.fi/portal/files/15671512/cosade2011.pdf},
  otherurl      = {https://researchportal.tuni.fi/en/publications/cache-timing-attacks-and-shared-contexts},
  url           = {https://researchportal.tuni.fi/files/15671512/cosade2011.pdf}
}

@InProceedings{DBLP:conf/fc/CoronKN00,
  author        = {Jean{-}S{\'{e}}bastien Coron and Paul C. Kocher and David Naccache},
  editor        = {Yair Frankel},
  title         = {Statistics and Secret Leakage},
  altbooktitle  = {Financial Cryptography},
  booktitle     = {Financial Cryptography, 4th International Conference, {FC} 2000 Anguilla, British West Indies, February 20-24, 2000, Proceedings},
  series        = {Lecture Notes in Computer Science},
  volume        = {1962},
  pages         = {157--173},
  publisher     = {Springer},
  year          = {2000},
  url           = {https://doi.org/10.1007/3-540-45472-1_12},
  ALTdoi        = {10.1007/3-540-45472-1_12},
  bibsource     = {dblp computer science bibliography, https://dblp.org}
}

@InProceedings{DBLP:conf/ches/AciicmezBG10,
  author        = {Onur Ac\i{}i\c{c}mez and Billy Bob Brumley and Philipp Grabher},
  editor        = {Stefan Mangard and Fran{\c{c}}ois{-}Xavier Standaert},
  title         = {New Results on Instruction Cache Attacks},
  altbooktitle  = {{CHES}},
  booktitle     = {Cryptographic Hardware and Embedded Systems, {CHES} 2010, 12th International Workshop, Santa Barbara, CA, USA, August 17-20, 2010. Proceedings},
  series        = {Lecture Notes in Computer Science},
  volume        = {6225},
  pages         = {110--124},
  publisher     = {Springer},
  year          = {2010},
  url           = {https://doi.org/10.1007/978-3-642-15031-9_8},
  ALTdoi        = {10.1007/978-3-642-15031-9_8}
}

@InProceedings{DBLP:conf/uss/YaromF14,
  author        = {Yuval Yarom and Katrina Falkner},
  title         = {{FLUSH+RELOAD:} {A} High Resolution, Low Noise, {L3} Cache Side-Channel Attack},
  altbooktitle  = {{USENIX} Sec.},
  booktitle     = {Proceedings of the 23rd {USENIX} Security Symposium, San Diego, CA, USA, August 20-22, 2014},
  pages         = {719--732},
  publisher     = {{USENIX} Association},
  year          = {2014},
  url           = {https://www.usenix.org/conference/usenixsecurity14/technical-sessions/presentation/yarom},
  isbn          = {978-1-931971-15-7}
}

@InProceedings{DBLP:conf/uss/LippGSMM16,
  author        = {Moritz Lipp and Daniel Gruss and Raphael Spreitzer and Cl{\'{e}}mentine Maurice and Stefan Mangard},
  editor        = {Thorsten Holz and Stefan Savage},
  title         = {{ARMageddon}: Cache Attacks on Mobile Devices},
  altbooktitle  = {{USENIX} Sec.},
  booktitle     = {25th {USENIX} Security Symposium, {USENIX} Security 16, Austin, TX, USA, August 10-12, 2016},
  pages         = {549--564},
  publisher     = {{USENIX} Association},
  year          = {2016},
  url           = {https://www.usenix.org/conference/usenixsecurity16/technical-sessions/presentation/lipp},
  bibsource     = {dblp computer science bibliography, https://dblp.org}
}

@InProceedings{DBLP:conf/uss/GreenLZIHE17,
  author        = {Marc Green and Leandro Rodrigues Lima and Andreas Zankl and Gorka Irazoqui and Johann Heyszl and Thomas Eisenbarth},
  editor        = {Engin Kirda and Thomas Ristenpart},
  title         = {{AutoLock}: Why Cache Attacks on {ARM} Are Harder Than You Think},
  altbooktitle  = {{USENIX} Sec.},
  booktitle     = {26th {USENIX} Security Symposium, {USENIX} Security 2017, Vancouver, BC, Canada, August 16-18, 2017},
  pages         = {1075--1091},
  publisher     = {{USENIX} Association},
  year          = {2017},
  url           = {https://www.usenix.org/conference/usenixsecurity17/technical-sessions/presentation/green},
  bibsource     = {dblp computer science bibliography, https://dblp.org}
}

@InProceedings{DBLP:conf/uss/GrasRBG18,
  author        = {Ben Gras and Kaveh Razavi and Herbert Bos and Cristiano Giuffrida},
  editor        = {William Enck and Adrienne Porter Felt},
  title         = {Translation Leak-aside Buffer: Defeating Cache Side-channel Protections with {TLB} Attacks},
  altbooktitle  = {{USENIX} Sec.},
  booktitle     = {27th {USENIX} Security Symposium, {USENIX} Security 2018, Baltimore, MD, USA, August 15-17, 2018},
  pages         = {955--972},
  publisher     = {{USENIX} Association},
  year          = {2018},
  url           = {https://www.usenix.org/conference/usenixsecurity18/presentation/gras},
  bibsource     = {dblp computer science bibliography, https://dblp.org}
}

@InProceedings{DBLP:conf/uss/PaccagnellaLF21,
  author        = {Riccardo Paccagnella and Licheng Luo and Christopher W. Fletcher},
  editor        = {Michael D. Bailey and Rachel Greenstadt},
  title         = {Lord of the Ring(s): Side Channel Attacks on the {CPU} On-Chip Ring Interconnect Are Practical},
  altbooktitle  = {{USENIX} Sec.},
  booktitle     = {30th {USENIX} Security Symposium, {USENIX} Security 2021, August 11-13, 2021},
  pages         = {645--662},
  publisher     = {{USENIX} Association},
  year          = {2021},
  url           = {https://www.usenix.org/conference/usenixsecurity21/presentation/paccagnella},
  bibsource     = {dblp computer science bibliography, https://dblp.org}
}

@InProceedings{DBLP:conf/uss/AldayaB22,
  author        = {Alejandro Cabrera Aldaya and Billy Bob Brumley},
  editor        = {Kevin R. B. Butler and Kurt Thomas},
  title         = {{HyperDegrade}: From {GHz} to {MHz} Effective {CPU} Frequencies},
  altbooktitle  = {{USENIX} Sec.},
  booktitle     = {31st {USENIX} Security Symposium, {USENIX} Security 2022, Boston, MA, USA, August 10-12, 2022},
  pages         = {2801--2818},
  publisher     = {{USENIX} Association},
  year          = {2022},
  url           = {https://www.usenix.org/conference/usenixsecurity22/presentation/aldaya},
  bibsource     = {dblp computer science bibliography, https://dblp.org}
}

@InProceedings{DBLP:conf/uss/YuDJKF23,
  author        = {Jiyong Yu and Aishani Dutta and Trent Jaeger and David Kohlbrenner and Christopher W. Fletcher},
  editor        = {Joseph A. Calandrino and Carmela Troncoso},
  title         = {Synchronization Storage Channels {(S2C)}: Timer-less Cache Side-Channel Attacks on the {Apple} {M1} via Hardware Synchronization Instructions},
  altbooktitle  = {{USENIX} Sec.},
  booktitle     = {32nd {USENIX} Security Symposium, {USENIX} Security 2023, Anaheim, CA, USA, August 9-11, 2023},
  pages         = {1973--1990},
  publisher     = {{USENIX} Association},
  year          = {2023},
  url           = {https://www.usenix.org/conference/usenixsecurity23/presentation/yu-jiyong},
  bibsource     = {dblp computer science bibliography, https://dblp.org}
}

@InProceedings{DBLP:conf/uss/WangPHSFK22,
  author        = {Yingchen Wang and Riccardo Paccagnella and Elizabeth Tang He and Hovav Shacham and Christopher W. Fletcher and David Kohlbrenner},
  editor        = {Kevin R. B. Butler and Kurt Thomas},
  title         = {Hertzbleed: Turning Power Side-Channel Attacks Into Remote Timing Attacks on x86},
  altbooktitle  = {{USENIX} Sec.},
  booktitle     = {31st {USENIX} Security Symposium, {USENIX} Security 2022, Boston, MA, USA, August 10-12, 2022},
  pages         = {679--697},
  publisher     = {{USENIX} Association},
  year          = {2022},
  url           = {https://www.usenix.org/conference/usenixsecurity22/presentation/wang-yingchen},
  bibsource     = {dblp computer science bibliography, https://dblp.org}
}

@InProceedings{DBLP:conf/acsac/Cronin0WC21,
  author        = {Patrick Cronin and Xing Gao and Haining Wang and Chase Cotton},
  title         = {An Exploration of {ARM} System-Level Cache and {GPU} Side Channels},
  altbooktitle  = {{ACSAC}},
  booktitle     = {{ACSAC} '21: Annual Computer Security Applications Conference, Virtual Event, USA, December 6-10, 2021},
  pages         = {784--795},
  publisher     = {{ACM}},
  year          = {2021},
  url           = {https://doi.org/10.1145/3485832.3485902},
  ALTdoi        = {10.1145/3485832.3485902},
  bibsource     = {dblp computer science bibliography, https://dblp.org}
}

@InProceedings{DBLP:conf/dac/KayaalpAPJ16,
  author        = {Mehmet Kayaalp and Nael B. Abu{-}Ghazaleh and Dmitry V. Ponomarev and Aamer Jaleel},
  title         = {A high-resolution side-channel attack on last-level cache},
  altbooktitle  = {{DAC}},
  booktitle     = {Proceedings of the 53rd Annual Design Automation Conference, {DAC} 2016, Austin, TX, USA, June 5-9, 2016},
  pages         = {72:1--72:6},
  publisher     = {{ACM}},
  year          = {2016},
  url           = {http://doi.acm.org/10.1145/2897937.2897962},
  ALTdoi        = {10.1145/2897937.2897962},
  bibsource     = {dblp computer science bibliography, https://dblp.org}
}

@InProceedings{DBLP:conf/dac/BarenghiP18,
  author        = {Alessandro Barenghi and Gerardo Pelosi},
  title         = {Side-channel security of superscalar {CPUs}: evaluating the impact of micro-architectural features},
  altbooktitle  = {{DAC}},
  booktitle     = {Proceedings of the 55th Annual Design Automation Conference, {DAC} 2018, San Francisco, CA, USA, June 24-29, 2018},
  pages         = {120:1--120:6},
  publisher     = {{ACM}},
  year          = {2018},
  url           = {https://doi.org/10.1145/3195970.3196112},
  ALTdoi        = {10.1145/3195970.3196112},
  bibsource     = {dblp computer science bibliography, https://dblp.org}
}

@InProceedings{DBLP:conf/sp/GullaschBK11,
  author        = {David Gullasch and Endre Bangerter and Stephan Krenn},
  title         = {Cache Games - Bringing Access-Based Cache Attacks on {AES} to Practice},
  altbooktitle  = {{IEEE} S{\&}P},
  booktitle     = {32nd {IEEE} Symposium on Security and Privacy, S{\&}P 2011, 22-25 May 2011, Berkeley, California, {USA}},
  pages         = {490--505},
  publisher     = {{IEEE} Computer Society},
  year          = {2011},
  url           = {https://doi.org/10.1109/SP.2011.22},
  ALTdoi        = {10.1109/SP.2011.22},
  bibsource     = {dblp computer science bibliography, https://dblp.org}
}

@InProceedings{DBLP:conf/sp/LiuYGHL15,
  author        = {Fangfei Liu and Yuval Yarom and Qian Ge and Gernot Heiser and Ruby B. Lee},
  title         = {Last-Level Cache Side-Channel Attacks are Practical},
  altbooktitle  = {{IEEE} S{\&}P},
  booktitle     = {2015 {IEEE} Symposium on Security and Privacy, {SP} 2015, San Jose, CA, USA, May 17-21, 2015},
  pages         = {605--622},
  publisher     = {{IEEE} Computer Society},
  year          = {2015},
  url           = {https://doi.org/10.1109/SP.2015.43},
  ALTdoi        = {10.1109/SP.2015.43},
  bibsource     = {dblp computer science bibliography, https://dblp.org}
}

@InProceedings{DBLP:conf/sp/AldayaBHGT19,
  author        = {Aldaya, Alejandro Cabrera and Brumley, Billy Bob and ul Hassan, Sohaib and Pereida Garc{\'{i}}a, Cesar and Tuveri, Nicola},
  title         = {Port Contention for Fun and Profit},
  altbooktitle  = {{IEEE} S{\&}P},
  booktitle     = {2019 {IEEE} Symposium on Security and Privacy, {SP} 2019, San Francisco, CA, USA, May 19-23, 2019},
  pages         = {870--887},
  publisher     = {{IEEE}},
  year          = {2019},
  url           = {https://doi.org/10.1109/SP.2019.00066},
  ALTdoi        = {10.1109/SP.2019.00066},
  bibsource     = {dblp computer science bibliography, https://dblp.org}
}

@InProceedings{DBLP:conf/sp/KocherHFGGHHLM019,
  author        = {Paul Kocher and Jann Horn and Anders Fogh and Daniel Genkin and Daniel Gruss and Werner Haas and Mike Hamburg and Moritz Lipp and Stefan Mangard and Thomas Prescher and Michael Schwarz and Yuval Yarom},
  title         = {Spectre Attacks: Exploiting Speculative Execution},
  altbooktitle  = {{IEEE} S{\&}P},
  booktitle     = {2019 {IEEE} Symposium on Security and Privacy, {SP} 2019, San Francisco, CA, USA, May 19-23, 2019},
  pages         = {1--19},
  publisher     = {{IEEE}},
  year          = {2019},
  url           = {https://doi.org/10.1109/SP.2019.00002},
  ALTdoi        = {10.1109/SP.2019.00002},
  bibsource     = {dblp computer science bibliography, https://dblp.org}
}

@InProceedings{DBLP:conf/sp/LippKOSECG21,
  author        = {Moritz Lipp and Andreas Kogler and David F. Oswald and Michael Schwarz and Catherine Easdon and Claudio Canella and Daniel Gruss},
  title         = {{PLATYPUS}: Software-based Power Side-Channel Attacks on x86},
  altbooktitle  = {{IEEE} S{\&}P},
  booktitle     = {42nd {IEEE} Symposium on Security and Privacy, {SP} 2021, San Francisco, CA, USA, 24-27 May 2021},
  pages         = {355--371},
  publisher     = {{IEEE}},
  year          = {2021},
  url           = {https://doi.org/10.1109/SP40001.2021.00063},
  ALTdoi        = {10.1109/SP40001.2021.00063},
  bibsource     = {dblp computer science bibliography, https://dblp.org}
}

@InProceedings{DBLP:conf/sp/GastJSSKFKG23,
  author        = {Stefan Gast and Jonas Juffinger and Martin Schwarzl and Gururaj Saileshwar and Andreas Kogler and Simone Franza and Markus K{\"{o}}stl and Daniel Gruss},
  title         = {{SQUIP}: Exploiting the Scheduler Queue Contention Side Channel},
  altbooktitle  = {{IEEE} S{\&}P},
  booktitle     = {44th {IEEE} Symposium on Security and Privacy, {SP} 2023, San Francisco, CA, USA, May 21-25, 2023},
  pages         = {2256--2272},
  publisher     = {{IEEE}},
  year          = {2023},
  url           = {https://doi.org/10.1109/SP46215.2023.10179368},
  ALTdoi        = {10.1109/SP46215.2023.10179368},
  bibsource     = {dblp computer science bibliography, https://dblp.org}
}

@InProceedings{DBLP:conf/isca/TullsenEL95,
  author        = {Dean M. Tullsen and Susan J. Eggers and Henry M. Levy},
  editor        = {David A. Patterson},
  title         = {Simultaneous Multithreading: Maximizing On-Chip Parallelism},
  altbooktitle  = {{ISCA}},
  booktitle     = {Proceedings of the 22nd Annual International Symposium on Computer Architecture, {ISCA} '95, Santa Margherita Ligure, Italy, June 22-24, 1995},
  pages         = {392--403},
  publisher     = {{ACM}},
  year          = {1995},
  url           = {https://doi.org/10.1145/223982.224449},
  ALTdoi        = {10.1145/223982.224449},
  bibsource     = {dblp computer science bibliography, https://dblp.org}
}

@InProceedings{DBLP:conf/hpca/KimHB22,
  author        = {Sowoong Kim and Myeonggyun Han and Woongki Baek},
  title         = {{DPrime+DAbort}: {A} High-Precision and Timer-Free Directory-Based Side-Channel Attack in Non-Inclusive Cache Hierarchies using {Intel} {TSX}},
  altbooktitle  = {{HPCA}},
  booktitle     = {{IEEE} International Symposium on High-Performance Computer Architecture, {HPCA} 2022, Seoul, South Korea, April 2-6, 2022},
  pages         = {67--81},
  publisher     = {{IEEE}},
  year          = {2022},
  url           = {https://doi.org/10.1109/HPCA53966.2022.00014},
  ALTdoi        = {10.1109/HPCA53966.2022.00014},
  bibsource     = {dblp computer science bibliography, https://dblp.org}
}

@InProceedings{dblp:journals/iacr/yaromgh16,
  author        = {Yuval Yarom and Daniel Genkin and Nadia Heninger},
  editor        = {Benedikt Gierlichs and Axel Y. Poschmann},
  title         = {{CacheBleed}: {A} Timing Attack on {OpenSSL} Constant Time {RSA}},
  altbooktitle  = {{CHES}},
  booktitle     = {Cryptographic Hardware and Embedded Systems - {CHES} 2016 - 18th International Conference, Santa Barbara, CA, USA, August 17-19, 2016, Proceedings},
  series        = {Lecture Notes in Computer Science},
  volume        = {9813},
  pages         = {346--367},
  publisher     = {Springer},
  year          = {2016},
  url           = {https://doi.org/10.1007/978-3-662-53140-2_17},
  ALTdoi        = {10.1007/978-3-662-53140-2_17}
}

@Article{dblp:journals/iacr/geych16,
  author        = {Qian Ge and Yuval Yarom and David Cock and Gernot Heiser},
  title         = {A survey of microarchitectural timing attacks and countermeasures on contemporary hardware},
  journal       = {J. Cryptographic Engineering},
  volume        = {8},
  number        = {1},
  pages         = {1--27},
  year          = {2018},
  url           = {https://doi.org/10.1007/s13389-016-0141-6},
  ALTdoi        = {10.1007/s13389-016-0141-6},
  bibsource     = {dblp computer science bibliography, https://dblp.org}
}

@InProceedings{dblp:journals/iacr/aciicmezsk06,
  author        = {Onur Ac\i{}i\c{c}mez and {\c{C}}etin Kaya Ko{\c{c}} and Jean{-}Pierre Seifert},
  editor        = {Masayuki Abe},
  title         = {Predicting Secret Keys Via Branch Prediction},
  altbooktitle  = {{CT-RSA}},
  booktitle     = {Topics in Cryptology - {CT-RSA} 2007, The Cryptographers' Track at the {RSA} Conference 2007, San Francisco, CA, USA, February 5-9, 2007, Proceedings},
  series        = {Lecture Notes in Computer Science},
  volume        = {4377},
  pages         = {225--242},
  publisher     = {Springer},
  year          = {2007},
  url           = {https://doi.org/10.1007/11967668_15},
  ALTdoi        = {10.1007/11967668_15}
}

@InProceedings{dblp:journals/corr/abs-1903-01843,
  author        = {Atri Bhattacharyya and Alexandra Sandulescu and Matthias Neugschwandtner and Alessandro Sorniotti and Babak Falsafi and Mathias Payer and Anil Kurmus},
  editor        = {Lorenzo Cavallaro and Johannes Kinder and XiaoFeng Wang and Jonathan Katz},
  title         = {{SMoTherSpectre}: Exploiting Speculative Execution through Port Contention},
  altbooktitle  = {{ACM} {CCS}},
  booktitle     = {Proceedings of the 2019 {ACM} {SIGSAC} Conference on Computer and Communications Security, {CCS} 2019, London, UK, November 11-15, 2019},
  pages         = {785--800},
  publisher     = {{ACM}},
  year          = {2019},
  url           = {https://doi.org/10.1145/3319535.3363194},
  ALTdoi        = {10.1145/3319535.3363194},
  bibsource     = {dblp computer science bibliography, https://dblp.org}
}

@InProceedings{templateattacks,
  author        = {Suresh Chari and Josyula R. Rao and Pankaj Rohatgi},
  editor        = {Burton S. Kaliski Jr. and {\c{C}}etin Kaya Ko{\c{c}} and Christof Paar},
  title         = {Template Attacks},
  altbooktitle  = {{CHES}},
  booktitle     = {Cryptographic Hardware and Embedded Systems - {CHES} 2002, 4th International Workshop, Redwood Shores, CA, USA, August 13-15, 2002, Revised Papers},
  series        = {Lecture Notes in Computer Science},
  volume        = {2523},
  pages         = {13--28},
  publisher     = {Springer},
  year          = {2002},
  url           = {https://doi.org/10.1007/3-540-36400-5_3},
  ALTdoi        = {10.1007/3-540-36400-5_3},
  bibsource     = {dblp computer science bibliography, http://dblp.org}
}

@InProceedings{nicv-paper,
  author        = {Shivam Bhasin and Jean{-}Luc Danger and Sylvain Guilley and Zakaria Najm},
  title         = {{NICV}: Normalized Inter-Class Variance for Detection of Side-Channel Leakage},
  altbooktitle  = {{EMC}},
  booktitle     = {International Symposium on Electromagnetic Compatibility, EMC 2014, Tokyo, Japan, May 12-16, 2014, Proceedings},
  year          = {2014},
  pages         = {310--313},
  url           = {https://ieeexplore.ieee.org/document/6997167}
}

@InProceedings{dblp:journals/iacr/osvikst05,
  author        = {Dag Arne Osvik and Adi Shamir and Eran Tromer},
  editor        = {David Pointcheval},
  title         = {Cache Attacks and Countermeasures: The Case of {AES}},
  altbooktitle  = {{CT-RSA}},
  booktitle     = {Topics in Cryptology - {CT-RSA} 2006, The Cryptographers' Track at the {RSA} Conference 2006, San Jose, CA, USA, February 13-17, 2006, Proceedings},
  series        = {Lecture Notes in Computer Science},
  volume        = {3860},
  pages         = {1--20},
  publisher     = {Springer},
  year          = {2006},
  url           = {https://doi.org/10.1007/11605805_1},
  ALTdoi        = {10.1007/11605805_1},
  bibsource     = {dblp computer science bibliography, https://dblp.org}
}

@InProceedings{welch,
  author        = {Gilbert Goodwill and Benjamin Jun and Josh Jaffe and Pankaj Rohatgi},
  title         = {A testing methodology for side-channel resistance validation},
  altbooktitle  = {{NIAT}},
  booktitle     = {Non-Invasive Attack Testing Workshop, NIAT 2011, Nara, Japan, September 26-27, 2011. Proceedings},
  publisher     = {{NIST}},
  year          = {2011},
  url           = {https://csrc.nist.gov/csrc/media/events/non-invasive-attack-testing-workshop/documents/08_goodwill.pdf}
}

@Article{dblp:journals/iacr/bernsteinbbcckkprt24,
  author        = {Daniel J. Bernstein and Karthikeyan Bhargavan and Shivam Bhasin and Anupam Chattopadhyay and Tee Kiah Chia and Matthias J. Kannwischer and Franziskus Kiefer and Thales B. Paiva and Prasanna Ravi and Goutam Tamvada},
  title         = {{KyberSlash}: Exploiting secret-dependent division timings in {Kyber} implementations},
  journal       = {{IACR} Trans. Cryptogr. Hardw. Embed. Syst.},
  volume        = {2025},
  number        = {2},
  pages         = {209--234},
  year          = {2025},
  url           = {https://doi.org/10.46586/tches.v2025.i2.209-234},
  ALTdoi        = {10.46586/TCHES.V2025.I2.209-234},
  bibsource     = {dblp computer science bibliography, https://dblp.org}
}

@Misc{bernstein2005cache,
  author        = {Daniel J. Bernstein},
  title         = {Cache-timing attacks on {AES}},
  year          = {2005},
  url           = {http://cr.yp.to/papers.html#cachetiming}
}

@InProceedings{dblp:journals/cacm/lippsgphhmkgyhs20,
  author        = {Moritz Lipp and Michael Schwarz and Daniel Gruss and Thomas Prescher and Werner Haas and Anders Fogh and Jann Horn and Stefan Mangard and Paul Kocher and Daniel Genkin and Yuval Yarom and Mike Hamburg},
  editor        = {William Enck and Adrienne Porter Felt},
  title         = {Meltdown: Reading Kernel Memory from User Space},
  altbooktitle  = {{USENIX} Sec.},
  booktitle     = {27th {USENIX} Security Symposium, {USENIX} Security 2018, Baltimore, MD, USA, August 15-17, 2018},
  pages         = {973--990},
  publisher     = {{USENIX} Association},
  year          = {2018},
  url           = {https://www.usenix.org/conference/usenixsecurity18/presentation/lipp},
  bibsource     = {dblp computer science bibliography, https://dblp.org}
}

@InProceedings{dblp:journals/iacr/aciicmez07,
  author        = {Onur Ac\i{}i\c{c}mez},
  editor        = {Peng Ning and Vijay Atluri},
  title         = {Yet another {MicroArchitectural} Attack: exploiting {I-Cache}},
  altbooktitle  = {{CSAW}},
  booktitle     = {Proceedings of the 2007 {ACM} workshop on Computer Security Architecture, {CSAW} 2007, Fairfax, VA, USA, November 2, 2007},
  pages         = {11--18},
  publisher     = {{ACM}},
  year          = {2007},
  url           = {https://doi.org/10.1145/1314466.1314469},
  ALTdoi        = {10.1145/1314466.1314469},
  bibsource     = {dblp computer science bibliography, https://dblp.org}
}

@misc{MIPSsheet, 
  author = {MIPS},
	title = {{MIPS} {1004K} coherent processing system datasheet},
	url = {https://s3-eu-west-1.amazonaws.com/downloads-mips/documents/MD00584-2B-1004K-DTS-01.20.pdf},
	year = {2011},
	note = {[Accessed 27-03-2025]},
}

@misc{MIPS-software-training,
	author = {MIPS},
	title = {{MIPS} software training},
	url = {https://training.mips.com/basic_mips/PDF/Caches.pdf},
	year = {2018},
	note = {[Accessed 27-03-2025]},
}

@misc{mips,
  author = {{MIPS}},
  title  = {Corporate Overview},
  url    = {https://mips.com/wp-content/uploads/2024/01/MIPS-Corporate-Overview-FINALdocx.pdf},
  year   = {2024},
  note   = {[Accessed 04-06-2026]},
}

\end{document}